\documentclass[final,5p,times,twocolumn]{art}
\usepackage{amssymb}
\journal{Physica E}

\begin{document}

\begin{frontmatter}

\title {A new type of localized fast moving electronic excitations in molecular chains}

\author{A.N. Korshunova\corref{cor1}}
\cortext[cor1]{Corresponding author, e-mail: alya@impb.psn.ru}
\author{V.D. Lakhno}

\address{Institute of Mathematical Problems of Biology, Russian Academy of
Sciences,  Pushchino, Moscow Region, 142290, Russia}%

\begin{abstract}
It is shown that in a Holstein molecular chain placed in a strong longitudinal electric field some new types of excitations can arise. This excitations can transfer a charge over large distance (more than 1000 nucleotide pairs) along the chain retaining approximately their shapes.
Excitations are formed only when a strong electric field either exists or quickly arises under especially preassigned conditions.
These excitations transfer a charge even in the case when Holstein polarons are practically immobile.
The results obtained are applied to synthetic homogeneous PolyG/PolyC \textit{DNA} duplexes.
They can be also provide the basis for explanation of famous H.W.Fink and C.Sch\"{o}nenberger experiment on long-range charge transfer in \textit{DNA}.
\end{abstract}

\end{frontmatter}

\section {Introduction}
A central problem of nanobioelectronics \cite{1_Lak_2008,2,123} is construction of molecular wires. During the past two decades a \textit{DNA} molecule which was demonstrated to possess conductivity in many experiments \cite{a3,3} has been considered a promising candidate for this role.
However this raises a principle problem of the mobility of charge carriers in homogeneous synthetic polynucleotide chains. Nanobioelectronics future largely depends on the decision of this issue.
The question of the charge motion in an electric field was considered also in \cite{c1} - \cite{c9}.
Experiments carried out so far cannot clarify the point unambiguously. Theoretical calculations for  $Poly G \big/ Poly C$ chains where  $G$ is guanine,  $C$ is cytosine yield the following.  In the case of "dry" chains the value of the hole mobility is  $\mu \sim 1cm^2/(V\cdot sec)$ \cite{4,5_Lafi_2003,6_Lafi_2005}. For "wet" chains, i.e. for chains in a solution where solvation effects are of great concern  \cite{7,8}, the mobility is several orders of magnitude lower \cite{8,9_Lafi_2012}. Notice that the terms "dry" and "wet" are conventional. Even in the case of dry \textit{DNA} there exists the hydration shell which stabilizes the double-helical conformation. This hydration shell only slightly influences on the mobility of carriers. As was shown in \cite{8} the solvation effect on carrier mobility is mainly conditioned by bulk polar solvent surrounded the molecule.

However, the notion of the mobility of charge carriers in itself is of little use for solving the problem of their conducting properties even for the case of homogeneous polynucleotide duplexes, since it applies to the situation of weak fields.
In typical experiments on charge transfer in \textit{DNA} the electric field intensity  $\mathcal{E}$ is $\mathcal{E} \sim 10^5\div10^6 V/cm$, i.e. it is not small.
In this case a charge can be transferred by mechanisms which are qualitatively different from those realized in weak fields.
This paper just deals with this problem.

\section {Numerical modeling of uniform motion in a molecular chain}
The conclusion that in weak fields the charge transfer is ineffective can be illustrated with a simple model of a Holstein polaron in a $Poly G \big/ Poly C$ chain described by the Hamiltonian:
\begin{eqnarray}\label{1}
   &H=-\sum_{n}^{N}\nu(|n\rangle\langle n-1|+|n\rangle\langle n+1|) + \sum_{n}^{N}\alpha q_n|n\rangle\langle n| \nonumber \\
   &+\sum_{n}^{N}M\dot{q}_n^2/2 + \sum_{n}^{N}kq_n^2/2 + \sum_{n}^{N}e\mathcal{E}an|n\rangle\langle n|,
\end{eqnarray}
where $\nu$ is a matrix element of the charge transition between neighboring sites (nucleotide pairs), $\alpha$ is a constant of the charge interaction with displacements $q_n$, $M$~is an effective mass of the site, $k$ is an elastic constant, $a=3.5\cdot10^{-8}cm$ is the distance between neighboring nucleotide pairs.

In the case under consideration a hole travels over guanine bases \cite{1_Lak_2008,2,3} and the process is described by the following motion equations:
\begin{eqnarray}
    i\hbar\dot{b}_n+\nu(b_{n-1}+b_{n+1})-\alpha q_n b_n-e\mathcal{E}anb_n=0,\label{2}\\
    M\ddot{q}_n+\gamma\dot{q}_n+kq_n+\alpha|b_n|^2=0\,,\label{3}
\end{eqnarray}
where $b_n$ is the amplitude of the probability of a charge occurrence at the $n$-th site, $\sum_n |b_n|^2=1$. Classical motion equations  (\ref{3}) include dissipation determined by the friction coefficient $\gamma$.

The parameters of a polynucleotide chain are taken to be $\nu=0.084eV$, $\alpha=0.13 eV/\mathring{A}$, $\gamma=6\cdot10^{-10}g/sec$, $M=10^{-21}g$, $k=0.062eV/\mathring{A}^2$ see  \cite{5_Lafi_2003,6_Lafi_2005,9_Lafi_2012}. As was mentioned earlier even the "dry" \textit{DNA} contains the absorbed water molecules and counterions. Because of the presence of absorbed molecules the introduced parameters need to be considered as an effective. It is important to notice, that the Holstein model under consideration for the case of small displacement coincides with Peyrard-Bishop-Dauxois model \cite{b1} (without stacking interaction) and for chosen parameters well reproduces the experimental results on photoinduced charge transfer in \textit{DNA} \cite{b2}.

As was shown in \cite{10_Lako_2010}, the excess charge injected into the chain (in the case under consideration this is a hole) in the absence of an applied external field  ($\mathcal{E}=0$), quickly forms a polaron state which is localized on two or three sites of the chain. As the field is now applied, for any field's value, the charge will remain localized practically on the same sites. This result is easy to verify by direct integration of motion equations (\ref{2}), (\ref{3})\footnote[1]{Notice that the unlimited growing of mobility in the limit of zero temperature \cite{5_Lafi_2003}, is connected with disregarding the formation of polaron states in the chain (Fig.1 in \cite{11_Lako_2011}).}.

\begin{figure}[t]
\resizebox{0.43\textwidth}{!}{\includegraphics{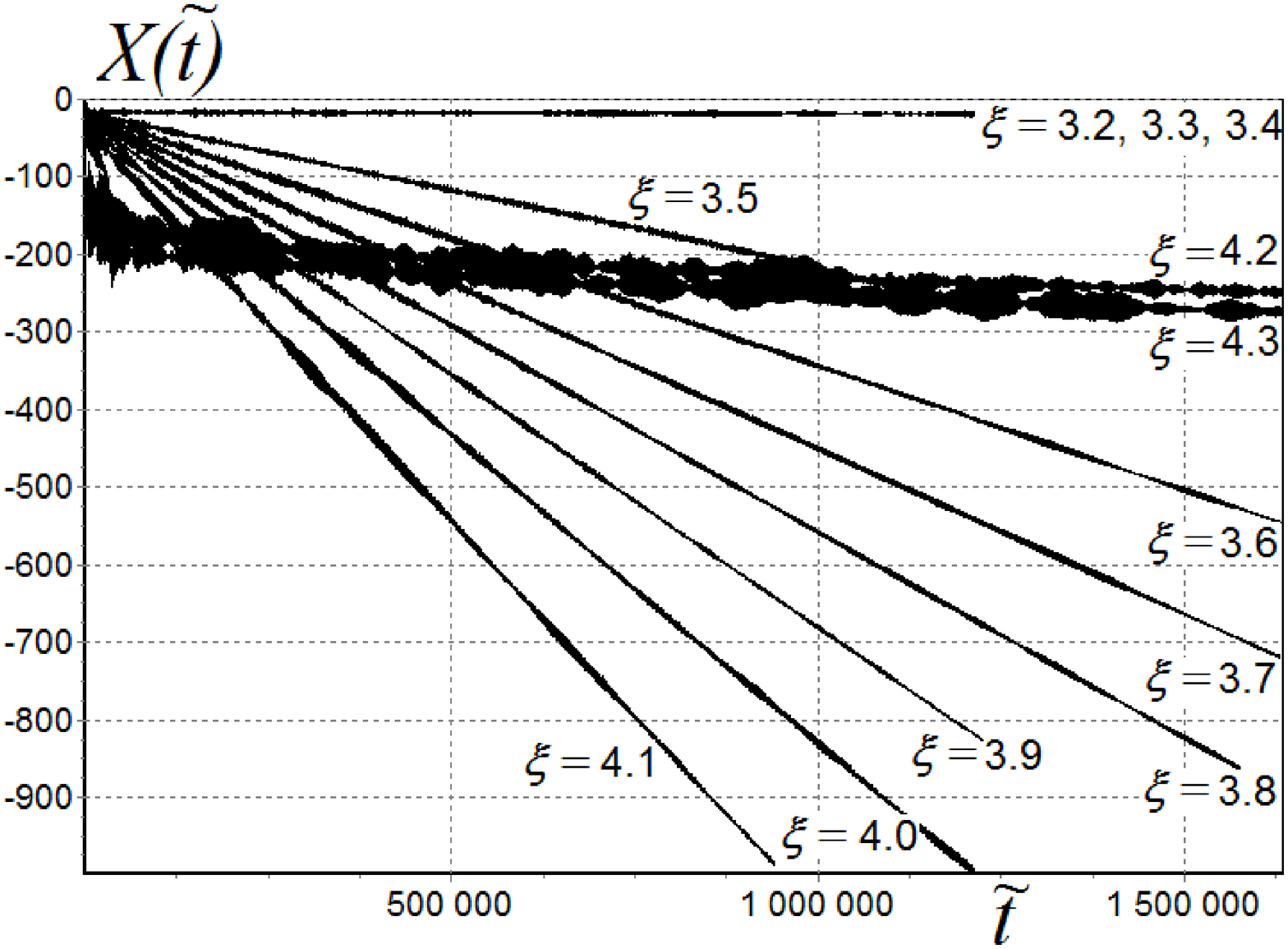}}
\caption{Graphs of the function $X(\widetilde{t})$ at the electric field intensity $\mathcal{E}=3.76\cdot10^{4}V/cm$ for (dimensionless) time $\widetilde{t}\approx1.6\cdot10^6$, $t=\widetilde{t}\cdot\tau, \tau=10^{-14}sec$. The values of the scaling coefficient $\xi$ are varying (see (\ref{4})).  The values of the chain parameters are: $\nu=0.084eV$, $\alpha=0.13 eV/\mathring{A}$, $\gamma=6\cdot10^{-10}g/sec$, $M=10^{-21}g$, $k=0.062eV/\mathring{A}^2$. The chain length is $N=1700$ sites.}\label{fig_0.02_X(t)}
\end{figure}

For the Holstein model, the problem of the uniform motion of a polaron in an electric field was studied in  \cite{11_Lako_2011}. As it shown in \cite{11_Lako_2011}, there exists a range of parameter values for which such a motion is possible (in particular, this is the case for a $Poly A \big/ Poly T$ chain\footnote[2]{As is shown in \cite{12_Lasu_2012}, for $Poly A / Poly T$ duplex a more complex situation is realized when the charge moves on both duplex chains, for which the considered model (1)\,-\,(3) becomes unapplicable.}). The parameter values given above do not fall into this region.

Noteworthy is that in \cite{11_Lako_2011} the ground state was considered as the initial one (i.e. polaron of an intermediate radius). Here we take the non-equilibrium state as the initial one - "scaled" ("stretched") solution of the equations (\ref{2}), (\ref{3}) in their continuum limit:
\begin{equation}\label{4}
 |b_n(0)|=\frac{1}{\sqrt{2\xi r}}\,\mathrm{ch}^{-1}\Bigl(\frac{n-n_0}{\xi r}\Bigr),
\end{equation}
 where $\xi$ is scaling coefficient, $r=4k\nu a\big/\alpha^2$.
The value of $n_0$ (the center of the initial state) in (\ref{4}) was chosen such that at the beginning of calculations the "initial polaron" \,be rather far away from the ends of the chain. Integration of (\ref{2})--(\ref{3}) was performed by the standard fourth-order Runge-Kutta method.

\begin{figure}[t]
\textbf{(a)}\resizebox{0.43\textwidth}{!}{\includegraphics{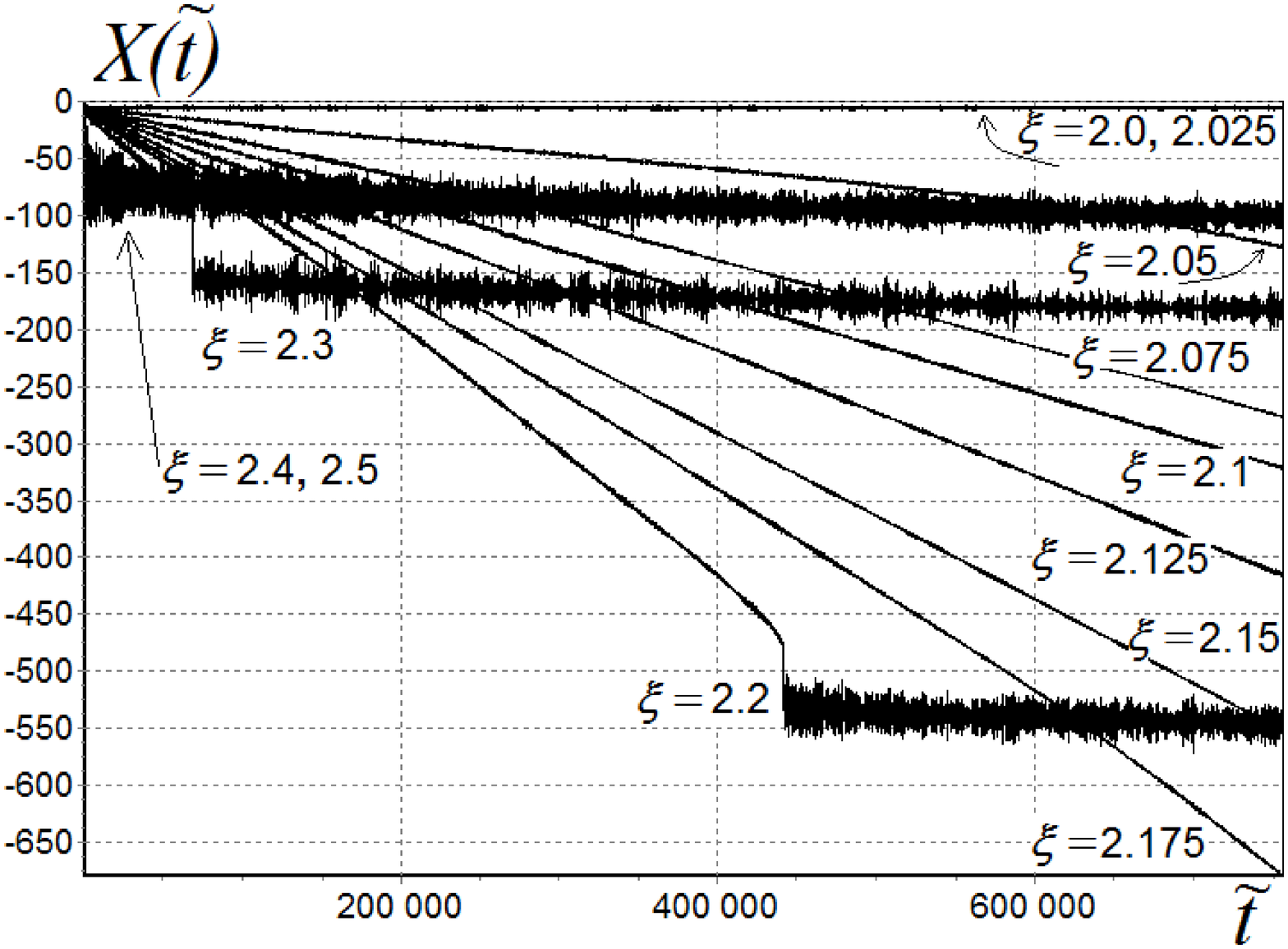}}
\textbf{(b)}\resizebox{0.21\textwidth}{!}{\includegraphics{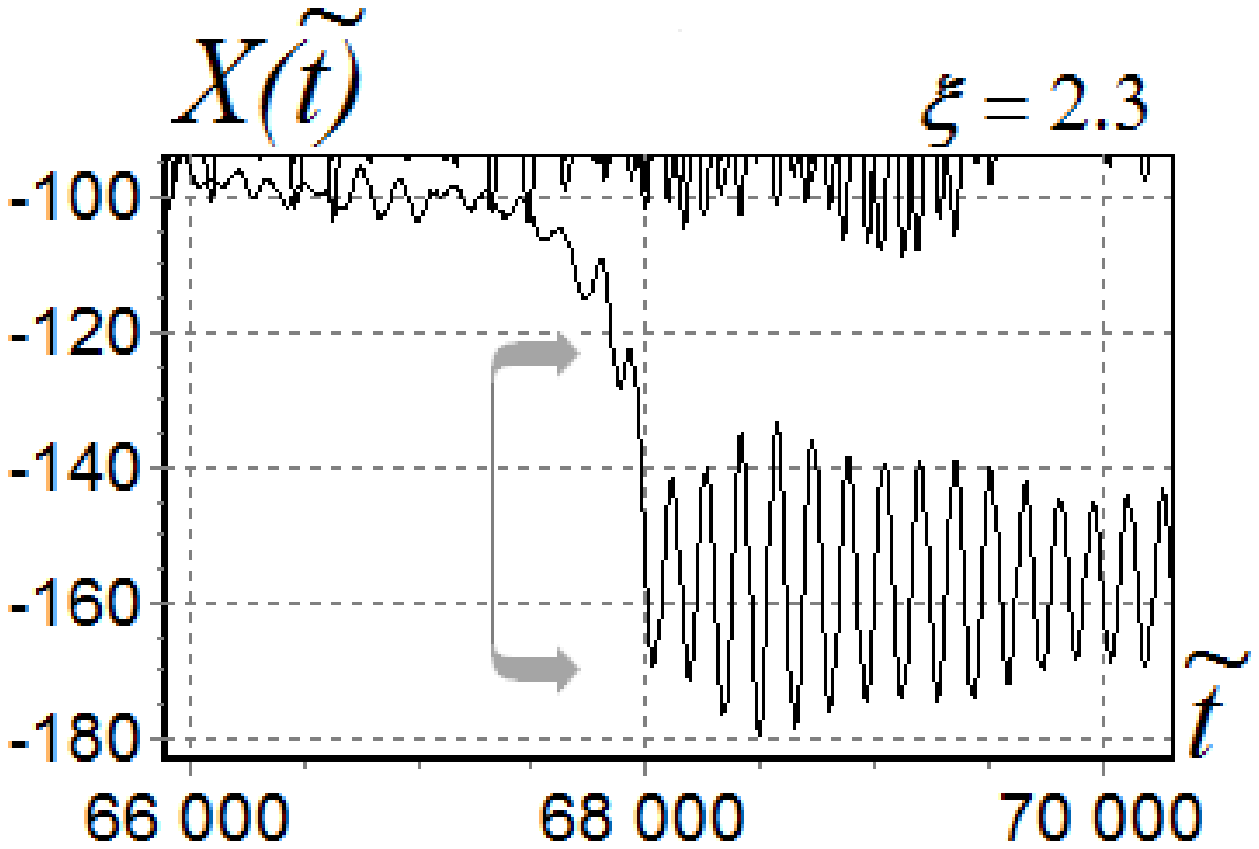}}
\textbf{(c)}\resizebox{0.21\textwidth}{!}{\includegraphics{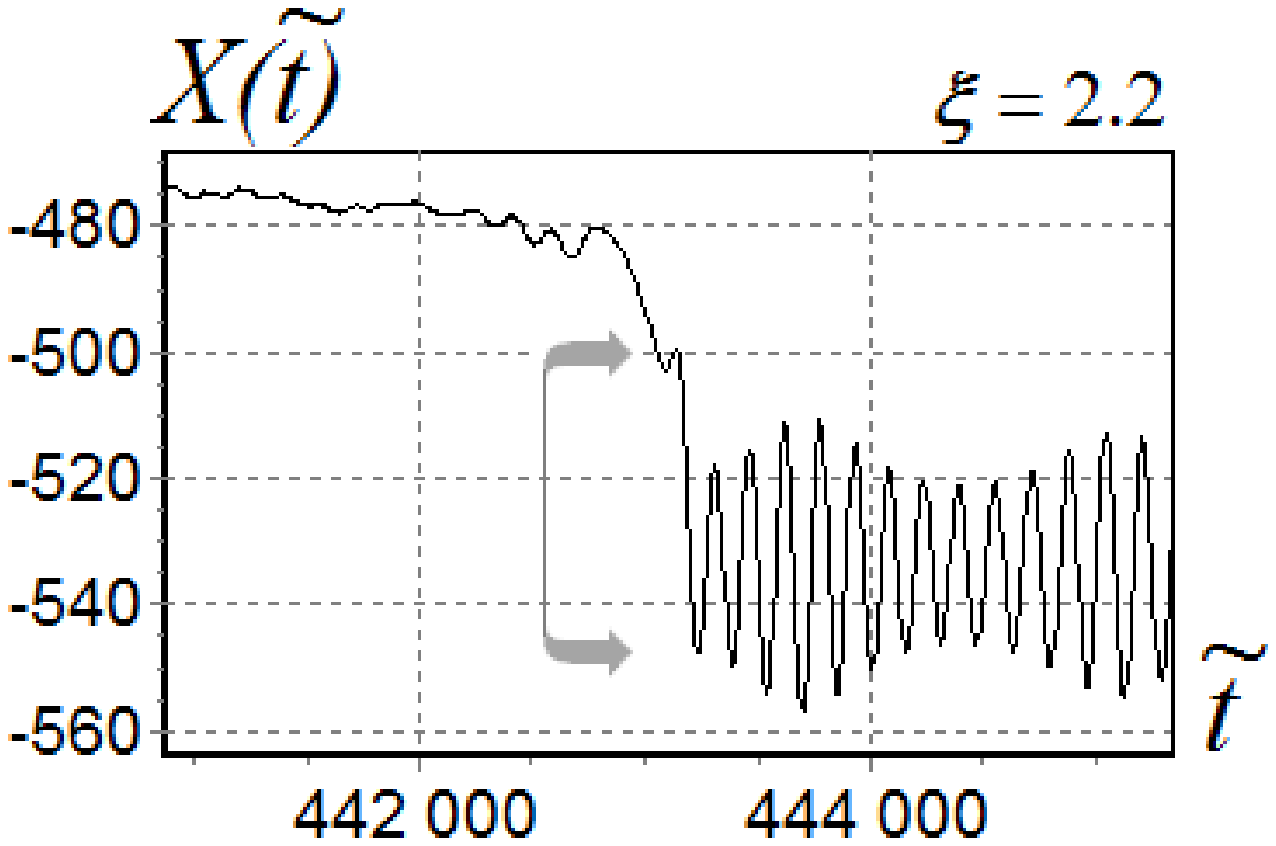}}
\caption{(a) -- Graphs of the function $X(\widetilde{t})$ for various values of the scaling coefficient $\xi$ for the above-cited values of the chain parameters and the electric field intensity $\mathcal{E}=7.52\cdot10^{4}V/cm$,  $\widetilde{t}\approx7.6\cdot10^5$. The chain length is $N=1700$ sites. (b),(c) -- are scaled up graphs of the function $X(\widetilde{t})$ for the moment of transformation from uniform motion to oscillatory one at $\xi=2.3$ and $\xi=2.2$ in Fig.\ref{fig_0.04_X(t)}(a).}\label{fig_0.04_X(t)}
\end{figure}

We will consider the values of the electric field intensity $\mathcal{E}$ lying on the interval $(0,\mathcal{E}_c)$, for which a polaron does not move under the initial condition (\ref{4}) when $\xi=1$.
$\mathcal{E}_c$ corresponds to the value of the field for which the polaron state falls apart and transforms into extended slowly moving state which experience the Bloch oscillation ($\mathcal{E}_c\approx1.3\cdot10^5V/cm$).

Fig.\ref{fig_0.02_X(t)} and Fig.\ref{fig_0.04_X(t)} present the graphs of the functions:
\begin{equation}\label{5}
 X(\widetilde{t})=\sum\nolimits_{n}{|\,b_n(\widetilde{t})|^2}\cdot (n-n_0)
\end{equation}
for various values of the parameter  $\xi$, where $X(\widetilde{t})$ has the meaning of the value of charge's displacement from its initial position along the chain during the time  $\widetilde{t}$ (in dimensional units the distance of charge transfer is $a\cdot|X(\widetilde{t})|$). The sites in the chain are numbered from left to right. Since we have chosen the value $\mathcal{E}>0$, the value of $n_0$ in (\ref{4}) was chosen at the right end of the chain. Excitations moves over the chain from right to left, therefore the values $X(\widetilde{t})<0$.

From figures \ref{fig_0.02_X(t)} and \ref{fig_0.04_X(t)} we can see that for each value of the electric field intensity $\mathcal{E}$, there exists a range of values of the scaling coefficients $\xi$, for which the graphs of $X(\widetilde{t})$ functions demonstrate on the whole a linear dependence on the time $\widetilde{t}$. Such linear dependence of $X(\widetilde{t})$ is the evidence of a uniform motion of the "scaled polaron" along the chain for very large distance - more then 1000 nucleotide pairs. This uniform motion is accompanied by low-amplitude Bloch oscillations which are clearly seen in Fig.\ref{fig_0.04_X(t)} (b) and (c).
These graphs also demonstrate that for the values of $\xi$  which are less than a certain critical value for a given $\mathcal{E}$, the "scaled polaron" either does not displace from its initial position at all, or moves over some sites and transforms into non-scaled standing state.
If the value of $\xi$  exceeds a certain allowed maximum value at which a uniform motion can take place for a certain preassigned value of  $\mathcal{E}$, the initial "scaled polaron" falls apart and this destroyed state moves further very slowly along the chain executing Bloch oscillations.

Fig.\ref{fig_0.04_X(t)}(a) clearly demonstrates that for large (for a given  $\mathcal{E}$) values of  $\xi$, a nonequilibrium polaron (in what follows excitation) moves at a constant velocity for rather a long time, but at a certain instant of time it falls apart and scarcely moves further. What will happen to a uniformly moving nonequilibrium excitation with the passage of much longer time, than in the presented graphs, we cannot say.

\section {Fast and superfast motions of localized excitations over long distances}

\begin{figure}[t]
\resizebox{0.43\textwidth}{!}{\includegraphics{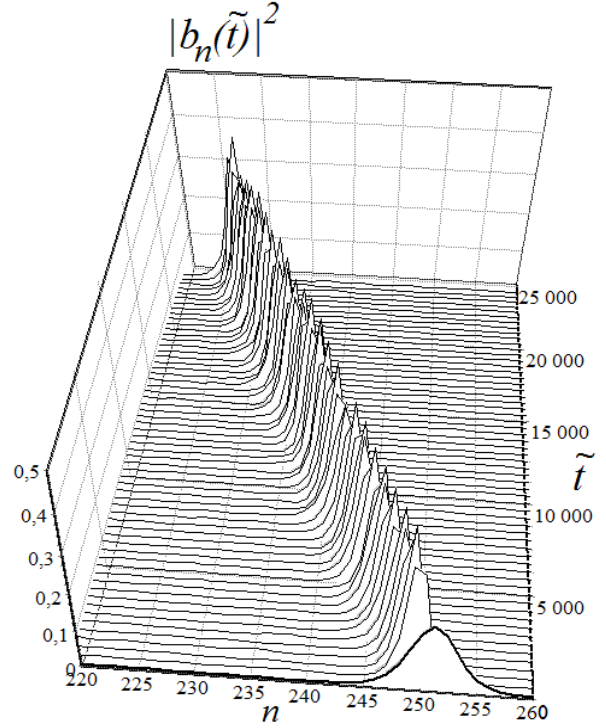}}
\caption{Graphs of the function $|\,b_n(\widetilde{t})|^2$ for the above-cited values of the chain parameters at $\xi=2.2$ and the electric field intensity $\mathcal{E}=7.52\cdot10^{4}V/cm$, $\widetilde{t}=0\rightarrow25000$. The chain length is $N=1700$ sites.}\label{fig_0.04_ksi_2.2_Bn}
\end{figure}

Now let us describe in more details a uniform motion of nonequilibrium polaron-type excitation from the initial nonequilibrium state for the above-cited parameter values of a $Poly G\big/Poly C$ chain. Fig.\ref{fig_0.04_ksi_2.2_Bn} presents a graph of the function $|\,b_n(\widetilde{t})|^2$ for the electric field intensity $\mathcal{E}=7.52\cdot10^{4}V/cm$ on the time interval $t=(0\div2.5\cdot10^{-10}sec)$ for a $Poly G\big/Poly C$ chain for the initial state (\ref{4}) with  $\xi=2.2$.
It is seen that the initial nonequilibrium state (\ref{4}) with $\xi=2.2$ and the characteristic size $L\approx9$ ($L=1\big/\sum_n |b_n|^4$)  has transformed into a nonequilibrium localized state of polaron type excitation with the characteristic size $L\approx5$  during a short time $t\approx5\cdot10^{-12}sec$  and, retaining its shape, has displaced along the chain by $25$ sites during the time $t=2.5\cdot10^{-10}sec$  which corresponds the  velocity of the excitation displacement $v\approx35m/sec$.
Notice that for each value of the electric field intensity  $\mathcal{E}$, the velocity of the uniform motion of the localized excitation increases as the scaling coefficient $\xi$ grows. Thus, for  $\xi=2.2$, the velocity of the excitation motion is approximately twice as great as that for  $\xi=2.1$ (see Fig.\ref{fig_0.04_X(t)}).
Hence, we can see that excitations move in a $Poly G\big/Poly C$ chain at different velocities for the same value of the electric field intensity  $\mathcal{E}$. It is also noteworthy that the maximum velocity of the uniform motion of an excitation is almost similar for various values of the electric field intensity  $\mathcal{E}$ and is approximately $v\approx35m/sec$  for a $Poly G\big/Poly C$ chain.

Fig.\ref{fig_0.04_X(t)} (b) and (c) shows two scaled up graphs from Fig.\ref{fig_0.04_X(t)}(a) on which we can see a transformation of the uniformly moving nonequilibrium excitation into the regime of oscillatory motion when a nonequilibrium excitation falls apart. On the short time interval when this transition occurs the velocity of the gravity center of the moving excitation sharply grows. In the regions marked by arrows in Fig. \ref{fig_0.04_X(t)} (b) and (c) the velocity of the gravity center is nearly  $v\approx 8600 m/sec$.

To observe the effect one should quickly apply an external field $\mathcal{E}$  or inject a charge in the initial nonequilibrium state into the chain to which an external field is already applied. If the electric field is applied not instantaneously at an initial time, but increases gradually to a certain predetermined value, the initial nonequilibrium state (\ref{4}) has time to transform to the relaxed one, i.e. the ground state which will not move in the electric field. Accordingly, if the field is removed as the excitation uniformly moves, the motion ceases and the excitation quickly transforms to the ground state. Such an "instantaneous"\, switch of the chain from the conducting state to the dielectric one can find an application in many nanobioelectronic devices.

\section{Conclusions}
In the foregoing we considered a uniform motion of a nonequilibrium polaron-type excitations from the initial scaled inverse hyperbolic cosine of the form (\ref{4}). Though the initial polaron of the form (\ref{4}) for $\xi=1$  is approximately a ground state (or a stationary solution of equations (\ref{2}), (\ref{3})), this does not make it an exclusive candidate for the function of the initial distribution at which the uniform motion in the chain starts. A uniform motion can be started from some other initial distributions similar in height and width to the distribution (\ref{4}) for various values of $\xi$  and satisfying the condition $\sum_n |b_n(0)|^2=1$. For example, if the initial state is taken to be the Gaussian function:
\begin{equation}\label{6}
 |b_n(0)|^2=\frac{1}{\sigma\sqrt{2\pi}}\cdot e^{-(n-n_0)^2/2\sigma^2},
\end{equation}
then for a given value of the electric field intensity  $\mathcal{E}$, we can find for (\ref{6}) a region of the dispersion $\sigma^2$  values for which a uniform motion exists. In this case the initial state of the form (\ref{6}) transforms into a nonequilibrium localized state with a smaller characteristic size which moves uniformly over the chain. This motion is similar to the above-described motion started from the scaled inverse hyperbolic cosine of the form (\ref{4}). Notice that the possibility of long-range charge transfer by nonequilibrium excitations seems to be rather general and earlier was considered for proteins in \cite{17,18,19}.

Up to now numerous experiments on charge transfer in \textit{DNA} have given contradictory results and have depended greatly on the length of a \textit{DNA} molecule, its nucleotide structure, availability of defects, the state of a solvent (if an experiment is done in a solution), the type of a substrate (if the molecule is on the surface of the substrate), the type of electrodes and, generally speaking, on the way of injecting the charge carriers into the molecule \cite{cn1} - \cite{cn5}.
In this paper we have shown that the results of an experiment will also greatly depend on the intensity of the applied external electric field and the initial distribution of the charge injected into the chain. The mechanism of charge transfer considered  opens up new  opportunities for development of \textit{DNA}-based electronic devices and molecular wires.

Finally we can say that we have considered the special case of homogeneous polynucleotide chain. Nevertheless we think, that the results obtained, may be rather general and may be applicable for heterogeneous nucleotide sequences. The reason for such assurance lies in sufficiently large characteristic size of moving excitations (about five nucleotide pairs for the electric field intensity $\mathcal{E}=7.52\cdot10^{4}V/cm$, moreover, characteristic size increases when decreasing the value of $\mathcal{E}$). When this occurs the chain for such excitations can be considered as homogeneous. If so the results obtained can explain the extremely long-range charge transfer (about 600$nm$) for $\lambda$ - \textit{DNA} in H.W.Fink and C.Sch\"{o}nenberger experiments \cite{20} which remains unexplained up to date.

The work has been done with the support from the RFBR, Project 13-07-00256.

\end{document}